\documentclass[prb,floats,superscriptaddress,showpacs]{revtex4}
\usepackage{amsmath}
\usepackage{graphicx}

\begin{document}
\date{\today}

\title{Non equilibrium inertial dynamics of colloidal systems}
\author{Umberto Marini Bettolo Marconi}
\affiliation{Dipartimento di Fisica, Via Madonna delle Carceri,
68032 Camerino (MC), Italy}
\affiliation{Departamento de Fisica Te\'orica de la Materia Condensada,
Universidad Autonoma de Madrid, E-28049 Madrid, Spain}
\author{Pedro Tarazona}
\affiliation{Departamento de Fisica Te\'orica de la Materia Condensada
and Instituto Nicol\'as Cabrera,
Universidad Autonoma de Madrid, E-28049 Madrid, Spain}
\begin{abstract}
We consider the properties of a one dimensional fluid of brownian inertial 
hard-core particles, whose microscopic dynamics 
is partially damped by a heat-bath.
Direct interactions among the particles are represented as binary, 
instantaneous elastic collisions.
Collisions with the heath bath are accounted for
by a Fokker-Planck collision operator, whereas direct collisions
among the particles are treated by a well known method
of kinetic theory, the Revised Enskog Theory.
By means of a time multiple time-scale method 
we derive the evolution equation for the average density.
Remarkably, for large values of the friction parameter and/or
of the mass of the particles 
we obtain the same equation as the one derived within the
dynamic density functional theory (DDF). In addition,
at moderate values of the friction constant,
the present method allows to study the
inertial effects not accounted for by DDF method.
Finally, a numerical test of these corrections is provided.  

\end{abstract}
\maketitle

\section{Introduction}
Over the last few years suspensions of interacting Brownian particles
have been the subject of vivid theoretical interest due to new accurate 
experiments probing their properties at nanoscale, down to the effects
of the correlation shells and layering structures in the density distribution.
There is a great variety of systems and problems of fundamental and applied
interest, including dense polymer solutions in good solvents~\cite{polymers},
the sedimentation of latex spheres, to produce materials
with engineered optical gaps~\cite{bolas}, the design of micro-fluidic 
devices, to handle colloidal 
particles~\cite{microfluidics1,microfluidics2,microfluidics3} 
or the crowding effects in the cellular cytoplasm~\cite{biophysics}. 
Whenever the systems may be considered to be at thermodynamical equilibrium,
the theoretical analysis of such structures may be efficiently done within
the Density Functional  formalism, with well tested approximations
to include the effects of the repulsive and attractive interactions
between the particles, although the inclusion of hydrodynamic (velocity
dependent) interactions is still an open challenge. 
The theoretical study of the dynamical properties 
of colloidal particles 
suspended in a solution of lighter particles is a much harder problem,  often 
studied through the Langevin approach to Brownian 
motion~\cite{Risken,Gardiner,Vankampen}, 
with the lighter particles represented by a bath providing a damping force, 
with friction constant $\gamma$, and a thermalizing stochastic noise. 

Two levels of description, both based on the Fokker-Planck
equation, can be employed to analyze Langevin model for Brownian motion. 
In the first, the so called Kramers equation~\cite{Kramers}
which governs the evolution of the joint probability distribution of position
and velocity, one keeps track both of velocities and positions
of the particles, whereas in the second, the Smoluchowski 
equation~\cite{Smoluchowski}, 
one considers only the evolution of the probability distribution of position.
In fact,
the velocity distribution relaxes in a time span of the order of
the inverse of the damping constant toward its equilibrium form 
and afterward remains stationary, so that
Kramers phase-space description becomes somehow redundant and one can 
restrict attention on the evolution of the spatial distribution,  
governed by Smoluchowski equation.  However,
the passage from the Kramers phase-space description to the Smoluchowski 
positional description 
requires the adiabatic elimination of the fast velocity variable.
Even for the simplest case of ideal non-interacting particles,
the correct procedure was understood only in the late seventies due to the
work of Wilemski~\cite{Wilemski} and Titulaer~\cite{Titulaer}. In particular,
Titulaer showed that a modified Smoluchowski equation
can be derived from Kramers equation by means of a systematic $\gamma^{-1}$ 
expansion of the Chapman-Enskog type. 
He obtained the corrections to the standard
Smoluchowski equation in terms of $\gamma$ for an arbitrary time independent
external potential.
 More recently, Bocquet et al.~\cite{Bocquet,Hansen} 
gave a pedagogical discussion of such a derivation using the multiple 
time-scale method~\cite{Nayfeh,Bender}.
The corrections to Smoluchowski equation for large, but finite, values of 
$\gamma$ represent the effects of the underlying inertial dynamics, over
the fully damped limit, in which  at any time the velocity of a particle,
averaged over the realization of the random noise, is  proportional 
to the external potential force, $\langle v (t)\rangle \sim F(x(t))$, 
with no inertial memory of the value of $\langle v(t')\rangle$ 
for $t'<t$.
 In the non-interacting case these corrections to the Smoluchowski equation
produce a gradient of the external force,
which determines a non uniform acceleration of the particle
and renormalizes the effective diffusion constant.

In this present paper, we are interested in the role played by the 
forces between the particles, in particular by those having a short range
character, such as the impulsive forces between hard spheres; which have been
usually neglected in previous studies. 
We want to answer the question whether inertial effects matter,
in the dynamics of a system of interacting colloidal particles, and which are
the corrections to the Smoluchowski equation in that case.
Interactions are expected to modify the motion of the particles by
restricting their trajectories, by inducing different accelerations, 
and correlating their velocities and positions. 
In the case of an over-damped dynamics, i.e. when $\gamma \to \infty$,
we presented~\cite{Tarazona}  
a dynamic equation that governs the probability density of
finding a particle in a given position. Starting from the Smoluchowski 
equation for the distribution function of the positions of $N$ particles,
we introduced a closure based on the assumption that the dynamical pair
correlations could be approximated by those of a reference equilibrium
system characterized by the same density profile as the non equilibrium 
system. The resulting self-consistent description for the average density
was encoded in a deterministic Dynamical Density Functional (DDF) 
equation:
\begin{equation}
\frac{\partial \rho(x,t)}{\partial t}=
D\nabla\Bigr [\nabla \rho(x,t)+\rho(x,t) \nabla( 
\frac{\delta \beta{\cal F}_{ni}[\rho]}{\delta \rho(x,t)} + \beta V_{ext}(x))\Bigr ]
\label{ddft}
\end{equation}
where $\beta {\cal F}_{ni}[\rho]$  is the non-ideal part of the
free energy functional, $\beta V_{ext}(x)$ is the external potential,
both in $\beta=(k_B T)^{-1}$ units~\cite{Archer}. The 
diffusion coefficient satisfies the Einstein relation
$D=\frac {k_B T}{ m \gamma}$, where $m$ is the mass of the colloidal
particles, $T$ the absolute temperature, $k_B$ the Boltzmann constant
and $\gamma$ the friction constant.
Notice that the essence of any DDF approach is to set an approximate
scheme in which the dynamic two-particle distribution function 
$\rho_2(x,x',t)$, required to include the interactions effects in 
the time derivative of $\rho(x,t)$, is taken as fully determined 
by the instantaneous value of that density distribution, as already 
done in earlier treatments like Enskog~\cite{Enskog} method and 
its revisions~\cite{Beijeren}.  The exact time evolution of $\rho(x,t)$ 
in interacting systems could only be obtained from the knowledge of the 
full previous history of the density distribution~\cite{Chan},
but from the practical point of view, the use of DDF approximations
seems to be well supported by the comparison of their predictions 
with Brownian Dynamics Simulations~\cite{Penna,Likos,Archer}.

Can we extend such a description to the case of systems, where the
dynamics is not over-damped? One would expect a richer dynamics as
compared to the purely diffusive dynamics of eq.(\ref{ddft}).
Does the momentum of the particles play a role? 
The one particle phase-space distribution function, $P[x,v,t]$,
is the natural candidate to replace the density $\rho(x,t)$ in this
extended description. Of course, the Boltzmann  equation for 
$P[x,v,t]$, which predates all non-equilibrium kinetic equations,
applies only to very dilute gases and does not incorporates the interaction
with an heat-bath. We shall consider both these aspects and show
that it is possible to derive eq.(\ref{ddft}) as the leading term of 
a $\gamma^{-1}$ expansion, starting from the full inertial dynamics. 
The leading corrections are also obtained as the next terms in the
expansion. 
In the present paper we investigate numerically and analytically the
problem in the simplified version of a one-dimensional colloidal fluid 
driven by an heat bath at fixed temperature. Although it may appear that 
the one dimensional model employed is not of direct practical relevance, our 
motivation derives not only from the great simplification of the resulting 
algebra and computer codes, 
but also from recent experimental work for colloidal
particles in very narrow channels~\cite{Cui}.

An outline of this article is as follows:
we open section II with a presentation of the microscopic model 
of inertial interacting particles subject to stochastic dynamics. 
We then introduce the evolution equation 
for the single particle phase-space distribution function
obtained by combining the effect of dissipative collisions with the 
heat-bath, which gives rise to a Kramers-Fokker-Planck contribution, 
with the effect of inter-particle collisions, described by an
Enskog collision term. At this stage, we separate the space dependence
from the velocity dependence of the phase space distribution functions
by using the eigenfunctions of the Fokker-Planck operator as basis
functions. As a result of such a projection procedure we obtain an infinite
non-linear system of coupled equations for the velocity moments of the 
phase distribution function.
 In section III by means of the
method of multiple-scales we construct a uniform expansion in the
inverse friction parameter and obtain the equation of evolution
for the particle density.
In section IV we explore the consequences of such an equation  with a simple
application and discuss its relation with the DDF equation.
Finally in section V we draw the conclusions.

\section{Enskog-Fokker-Planck equation}
Let us consider a system of heavy particles suspended in a solution of lighter
particles. Due to their smaller mass, the solvent particles perform rapid
motions so that their influence on the heavy particles
can be described by a stochastic force. As a
result of such elimination of microscopic degrees of freedom
one can represent the motion  heavy particles
by  means of stochastic Langevin dynamics. 
Here we consider a system of $N$ particles moving in one dimension,
under the action of an external force $f_e(x)$ and  
interacting elastically with a pair potential energy $U(x-x')$.
The equations of motions are:
\begin{eqnarray}
&\frac{d }{dt} x_i &= v_i \\
& m \frac{d v_i}{dt}& = -m \gamma v_i +f_e(x_i)-\sum_{j\neq i} 
\frac{\partial U(x_i-x_j)}{\partial x_i} +\xi_i(t),
\label{kramers}
\end{eqnarray}
including the effects of the solvent with the linear friction coefficient
$\gamma$, and the 
stochastic white noise with zero average and  correlation 
\begin{equation}
\langle \xi_i(t)\xi_j(s) \rangle  = 2 \gamma m k_B T\delta_{ij} \delta(t-s)\;,
\end{equation} 
$T$ is the ``heat-bath temperature'' and $\langle \cdot \rangle$ indicates 
the average over a statistical ensemble of realizations~\cite{Pagnani,Cecprl}.
The elimination in (\ref{kramers}) of the rapid bath variables $\xi_i(t)$
leads to the
Fokker-Planck equation~\cite{Risken,Gardiner}, in terms of the 
probability distribution function, $P(x,v,t)$ for the position and velocity
variables,
\begin{eqnarray}
\frac{\partial}{\partial t}P(x,v,t)+
\Bigr [v \frac{\partial}{\partial x}+\frac{f_e(x)}{m}
\frac{\partial}{\partial v} \Bigr] P(x,v,t)=
\gamma \Bigr[\frac{\partial}{\partial v} v  
+\frac{T}{m}\frac{\partial^2}{\partial v^2} \Bigr] P(x,v,t)+
{\cal C}[x,v,t,P_2].
\label{fokker}
\end{eqnarray} 
The l.h.s is the Liouville operator for the ideal gas, under
the external force $f_e(x)$, the first term 
in the r.h.s. represents the heat-bath as the  
standard Fokker-Planck collision operator,
and the last term represents the effect of the interactions among the 
particles, 
as a generic collision operator,
\begin{eqnarray}
{\cal C}[x,v,t,P_2]=
\frac{1}{m}\frac{\partial}{\partial v} \int dx'\int dv' 
 \frac{\partial U(x-x')}{\partial x}
P_{2}(x,v,x',v',t), 
\label{Ke} 
\end{eqnarray} 
This operator ${\cal C}$ 
satisfies the first equation of the
Bogoliubov-Born-Green-Kirkwood-Yvon (BBGKY) hierarchy~\cite{Bogoliubov}, which
connects the evolution of the $n$-particle distribution function,
to the distribution function for  $(n+1)$ particles. 
For interacting particles, the evolution equation~(\ref{fokker}) 
for the one-particle distribution 
function $P(x,v,t)$ depends on the two-particle distribution 
$P_{2}(x,v,x',v',t)$, and some approximate closure is required
to obtain a workable scheme. 

 Whenever $U(x-x')$ is a smooth function of the particle
separation, like for the ultra-soft repulsive potentials used to model
the steric repulsion between polymers~\cite{polymers}, or in the 
long range attractive interactions from dispersion or screened ionic
forces, we may follow a mean-field approximation $P_{2}(x,v,x',v',t)\approx
P(x,v,t) P(x',v',t)$, which reduces (\ref{fokker}) to a partial differential
equation for the one-particle distribution~\cite{Stell,Sobrino},
and the 
effects of the particle interactions may be 
directly integrated over velocities,
with the density distribution
$
\rho(x,t)=\int dv P(x,v,t);
$
and included as a molecular field,
$
f_m(x,t)=-\int dx'\frac{\partial U(x-x')}{\partial x}\rho(x',t),
$
to be added to $f_e(x)$ in the l.h.s. of (\ref{fokker}), 
as a self-consistent, $\rho(x,t)$ dependent, force field.

On the other hand, sharp repulsive contributions between the particles
cannot be included as a molecular field, since they imply very strong
correlations between the relative position $(x-x')$ and the relative
velocity $(v-v')$ over the range of the repulsive
force;  so that $P_{2}(x,v,x',v',t)$ goes sharply to zero when $x-x'$ goes
into the repulsive core. For hard-rod particles, of 
length $\sigma$, there is an
infinite force acting on an infinitesimal range around $x-x'=\pm \sigma$, and
the collision operator ${\cal C}[x,v,t,P_2]$
is exactly represented~\cite{Ernst} by the following operator 
\begin{equation}
K_E[x_1,v_1,t] = 
\sum_{s=\pm 1} 
\int dx_2  \int dv_2   
\Theta(v_{12}s)(v_{12}s)
\bigl[\delta(x_{12}-s\sigma)b_{12}-\delta(x_{12}+s\sigma)\bigl]
P_2[x_1,v_1,x_2,v_2,t],
\label{jcollis0}
\end{equation}
where $\Theta$ is the Heaviside function, 
$v_{12}=(v_1-v_2)$, $x_{12}=(x_1-x_2)$  and
$b_{12}$ is the scattering operator defined for arbitrary function
$X(v_1,v_2)$ by
\begin{equation}
b_{12}X(v_1,v_2)=X(v_1',v_2'),
\label{scat}
\end{equation} 
which for hard-rods swaps the velocities, $b_{12}X(v_1,v_2)=X(v_2,v_1)$,
thus generating a correlation between relative
position and relative velocity.
The representation (\ref{jcollis0}) formally integrates (\ref{Ke}) 
over the instant 
of collision, and substitutes the direct effect of the force by the
change from the pre-collisional to the post-collisional velocities. 

A standard approximation of the collision term (\ref{jcollis0}) is 
to assume that
atoms are uncorrelated immediately prior to collision,
which is the essence of Boltzmann's ``Molecular chaos hypothesis'',
but are correlated after they collide, because the collision 
itself generates correlations~\cite{Lutsko}.
The revised Enskog theory
(RET), developed by van Beijeren and Ernst~\cite{Beijeren},
truncates the infinite BBGKY hierarchy by factorizing
\begin{equation}
P_2(x_1,v_1,x_2,v_2,t)=g_2[x_1,x_2;\rho]P(x_1,v_1,t)P(x_2,v_2,t),
\label{enskog0}
\end{equation}
The spatial pair distribution function, $g_2[x_1,x_2;\rho]$,
reflects the local positional correlations in the fluid.
For particles with both
short-range repulsions and long-range tails, the simplest approximation
would be to split the generic collision operator $K_E$ in (\ref{Ke})
into a molecular field representation of the soft interactions and an
effective hard-rods description (\ref{jcollis0}) of the core repulsion,
following the usual treatment for equilibrium properties, which goes back
to van der Waals, and is still the most used scheme within the Density
Functional Formalism.    

In the case of one dimensional elastic hard-rods the RET 
provides the following expression for the collision integral: 
\begin{eqnarray}
&&K_E[x_1,v_1,t] =\sum_{s=\pm 1} \int dv_2\Theta(v_{12}s) 
(v_{12}s) \times \\ \nonumber
&&\bigl\{g_2[x_1,x_1-s\sigma;n]
P[x_1,v_1',t]P[x_1-s\sigma,v_2',t]-
g_2[x_1,x_1+s\sigma;n] P[x_1,v_1,t]P[x_1+s\sigma,v_2,t]\bigl\}
\label{jcollis1}
\end{eqnarray}
Whereas in  
Enskog's formulation
the pair correlation function at contact was assumed to be that 
of an equilibrium fluid evaluated 
at the local density at some point in between the colliding atoms,
in the RET instead the contact value of $g_2$ is assumed: i) to be a 
non-local equilibrium functional of the local density,
ii) to depend on time only through the
density $\rho(x,t)$ and iii) to have the same form 
as in a nonuniform equilibrium state whose density profile is $\rho(x,t)$.
Fortunately, in the case of a one dimensional hard-rod system the
exact expression for the equilibrium pair correlation at contact is
known given any arbitrary equilibrium density profile and
reads~\cite{Percus}:
\begin{equation}
g_2[x \pm \sigma;\rho]=\frac{1}{1-\eta (x \pm\frac{\sigma}{2})}
\qquad .
\label{g2}
\end{equation}
The density dependence occurs entirely via the
local packing fraction
$\eta(x,t)=\int_{x-\sigma/2}^{x+\sigma/2}dx' \rho(x',t)$.

At this stage it is convenient to introduce 
the following dimensionless variables: 
\begin{equation}
\tau\equiv t\frac{v_T}{\sigma}, \qquad V\equiv\frac{v}{v_T}, 
\qquad X\equiv\frac{x}{\sigma},\qquad \Gamma=\gamma\frac{\sigma}{v_T}
\label{adim1}
\end{equation}

\begin{equation}
F_e(X)\equiv\frac{\sigma f_e(x)}{m v_T^2},\qquad 
F_m(X,\tau)\equiv\frac{\sigma f_m(x,t)}{m v_T^2}
\label{adim2}
\end{equation}
\begin{equation}
\tilde P(X,V,\tau)\equiv \sigma v_T P(x,v,t), \qquad
K(X,V,\tau)\equiv \sigma^2 K_E(x,v,t)
\label{adim3}
\end{equation}
where $v_T=\sqrt{k_B T/m}$.


In situations where $\Gamma>>1$ particles lose memory of their initial
velocities after a time span which is of the order of the inverse
of the friction coefficient $\gamma$ so that the velocity distribution 
soon becomes a Maxwellian.
On the other hand, during the same interval
the coordinates of the particles suffer a negligible change, as one can
see comparing the product of the thermal velocity $v_T$ by $\gamma^{-1}$
with the typical molecular size $\sigma$. 
In this limit the Smoluchowski description of a system of non interacting
particles, which takes into account
only the configurational degrees of freedom, turns out to be adequate.
However, for intermediate values of $\Gamma$ inertial effects may
come into play. The question is how do we recover a description similar
to that provided by the DDF approach starting from a phase description?
On physical grounds one could directly neglect the inertial term in equation
(\ref{kramers}) and consider only the evolution 
of the position distribution, as the DDF does, but such an
approach does not give a clue on how the the inertial effects
can modify the dynamics.

Accordingly, Kramers' evolution equation for 
the phase space distribution function
can be rewritten with the help of relations (\ref{adim1}-\ref{adim2}) 
and with the definition of effective field $F(X,\tau)=F_e(X)+F_m(X,\tau)$ 
as:
\begin{equation}
\frac{1}{\Gamma}\frac{\partial \tilde P(X,V,\tau)}{\partial \tau}
=L_{FP}\tilde P(X,V,\tau)
-\frac{1}{\Gamma}V \frac{\partial }{\partial X} \tilde P(X,V,\tau) 
-\frac{1}{\Gamma}F(X,\tau) \frac{\partial }{\partial V} \tilde P(X,V,\tau)+
\frac{1}{\Gamma} K(X,V,\tau)
\label{kramers0}
\end{equation}
having introduced the ``Fokker-Planck'' operator
$
L_{FP}\tilde P(X,V,\tau) =\frac{\partial}{\partial V}\Bigl[
\frac{\partial }{\partial V }+V\Bigl]  \tilde P(X,V,\tau)
$,
whose eigenfunctions $H_{\nu}(V)
H_{\nu}(V)\equiv \frac{1}{\sqrt{2\pi}}
(-1)^{\nu} \frac{\partial^{\nu}}{\partial V^{\nu}} \exp(-\frac{1}{2}V^2)
$
have non positive integer eigenvalues $\nu=0,-1,-2,..$.
Solutions of eq. (\ref{kramers0}),
where position and velocity dependence of the distribution function 
are separated, can be written as:
\begin{equation}
\tilde P(X,V,\tau) \equiv \sum_{\nu=0}^{\infty}\phi_{\nu}(X,\tau) H_{\nu}(V).
\label{expansion}
\end{equation}
Moreover,
by multiplying $K(X,V,\tau)$ by $\frac{1}{n!}H_n(V)/H_0(V)$
and integrating with respect to $V$, one represents the collision term as
\begin{equation}
\tilde K(X,V,\tau) \equiv \sum_{\nu=0}^{\infty}C_{\nu}(X,\tau) H_{\nu}(V).
\label{expansion3}
\end{equation}

After  substituting (\ref{expansion}) and (\ref{expansion3}) into 
eq.~(\ref{kramers0}) we find
\begin{eqnarray}
\sum_{\nu}\Bigl[&&\frac{\partial \phi_{\nu}(X,\tau)}{\partial \tau}
+\Gamma \nu\phi_{\nu}(X,\tau)- C_\nu(X,\tau)\Bigl] H_\nu(V)+
\\\nonumber
&&\Bigl[\frac{\partial \phi_\nu(X,\tau)}{\partial X}
-F(X)\phi_\nu(X,\tau) \Bigl]H_{\nu+1}(V)+\nu \frac{\partial \phi_\nu(X,\tau)}
{\partial X}(\delta_{\nu,0}-1)H_{\nu-1}(V) =0
\label{brinkmanhiera}
\end{eqnarray}
Finally, by
equating the coefficients of the same basis functions, $H_{\nu}$,
we obtain an infinite  hierarchy of equations which differs from
standard Brinkman's expansion~\cite{Brinkman} by the presence
of collision terms.

\subsection{Physical interpretation of the expansion}

Before considering in detail the method of solution, we digress on the
physical interpretation of our equations.
By identifying $\phi_0(X,\tau)$ with the dimensionless
particle density, $n=\rho\sigma$, $\phi_1(X,\tau)$
with the momentum flow density, $J_v$, 
$\phi_2=E_k-n/2$ 
with the deviation from the thermalized value of the kinetic energy,
$E_k$ being the 
kinetic energy density, expressed in reduced units,
we can rewrite  the first three equations:
\begin{equation}
\frac{\partial n(X,\tau)}{\partial \tau}
=-\frac{\partial J_v(X,\tau)}{\partial X}
\label{hiera0b}
\end{equation}
\begin{equation}
\frac{\partial J_v(X,\tau)}{\partial \tau}
=-\Gamma J_v(X,\tau)+F(X,\tau)n(X,\tau)
-2\frac{\partial E_k(X,\tau)}{\partial X}+ C_1(X,\tau)
\label{hiera1b}
\end{equation}
\begin{equation}
\frac{\partial E_k(X,\tau)}{\partial \tau}=
-2\Gamma \Bigl [E_k(X,\tau)-\frac{1}{2}n(X,\tau)\Bigl]
-\frac{\partial J_k(X,\tau)}{\partial X}+
F(X,\tau) J_v(X,\tau)+ C_2(X,\tau)
\label{hiera2b}
\end{equation}
where the kinetic energy flow
is defined as $J_k\equiv \int dV V^3 \tilde P(X,V,\tau)$.

Using the result derived in Appendix B we can express the 
coefficients $C_n(X,\tau)$ as divergences.
First we
introduce~\cite{Barreiro} the kinetic pressure $\Pi_k=2 E_k$
and second identify the collisional contributions to the pressure and 
to the energy current via
\begin{equation}
\frac{\partial \Pi_c(X,\tau)}{\partial X}=
- C_1(X,\tau) \qquad ,
\frac{\partial J_c(X,\tau)}{\partial X}=
- C_2(X,\tau).
\label{c2b}
\end{equation}
We arrive at
\begin{equation}
\frac{\partial J_v(X,\tau)}{\partial \tau}
=-\Gamma J_v(X,\tau)+F(X,\tau)n(X,\tau)
-\frac{\partial [\Pi_k(X,\tau)+\Pi_c(X,\tau)] }{\partial X}
\label{hiera1bb}
\end{equation}
and
\begin{equation}
\frac{\partial E_k(X,\tau)}{\partial \tau}=
-2\Gamma \Bigl [E_k(X,\tau)-\frac{1}{2}n(X,\tau)\Bigl]
+F(X,\tau) J_v(X,\tau)
-\frac{\partial [J_k(X,\tau)+J_c(X,\tau)]}{\partial X}
\end{equation}
where the presence of the source 
term $n(X,\tau)/2$ maintains the fluid at
constant temperature.
Notice that properties (\ref{c2b})
are consequences of the local conservation of 
momentum and energy during the collisional process.
In one-dimensional elastic systems 
in addition to mass, impulse and energy
all higher moments of the
velocity distribution are conserved quantities under collisions, because
$K(X,V,\tau)$ is a divergence.

If the hierarchy of moment equations is truncated,
by supplementing the constitutive equations, 
one recovers the analogue of hydrodynamic equations with
dissipation. 
 We also remark that in a uniform bulk system 
the collisional contribution to the pressure coincides with the pressure
excess over the ideal gas pressure, since
$\Pi_c(X,\tau)=n^2/(1-n)$  , having used eq.(\ref{diverg})
and the  contact value, $g_2^b=1/(1-n)$,  of the bulk pair correlation.

\subsection{Exact solution for the free ideal gas.}

We illustrate the nature of the solutions
by a simple example,
namely the free-expansion of a system where the collisional terms 
and the molecular force field are dropped.
Let us remark, that even in that simple case, the time evolution
of the inhomogeneous ideal gas, is not well described by any
simple truncation of the hierarchy, for instance setting $\phi_3(X,\tau)=0$
in order to obtain a closed system of equations for the first three
weight functions. The  exact eigenfunctions of Kramers' equation are 
known, and they can be expressed as infinite series of the form
\begin{equation}
\tilde P^{(\mu)}(X,V,\tau)= \exp(- \mu \Gamma \tau) 
\exp\left[ - \frac{A_{+}}{\Gamma} 
\frac{\partial}{\partial x} \right] \left(1+ \frac{A_{-}}{\Gamma}
\frac{\partial}{\partial x} \right)^\mu H_{\mu}(V) \phi_{0}^{(\mu)}(X,\tau),
\end{equation}
where $A_{+}$ and $A_{-}$ are the raising and lowering operators on the 
FP velocity eigenfunctions, respectively,, 
$A_{\pm}H_{\nu}(V)=H_{\nu \pm 1}(V)$.
The functions $\phi_{0}^{(\mu)}(X,\tau)$, which fully define $\tilde P^{(\mu)}(X,V,\tau)$
, may be any generic solutions of the diffusion equation
\begin{equation}
\frac{\partial}{\partial \tau}\phi_{0}^{(\mu)}(X,\tau)=
\frac{1}{\Gamma}\frac{\partial^2}{\partial X^2}\phi_{0}^{(\mu)}(X,\tau)
\label{diff}
\end{equation}
which produces the time dependence to be scaled as $\tau_1\equiv \tau/\Gamma$.
Therefore, for $\Gamma \gg 1$ there is a clear separation 
between the {\it fast}
time dependence of the exponential 
decay $exp(- \mu \Gamma \tau)$ and the
{\it slow} dependence of the function $\phi_{0}^{(\mu)}(X,\tau)$. 
The eigenfunction associated with $\mu=0$ has the explicit form
\begin{equation}
\tilde P^{(0)}(X,V,\tau)= exp\left[ - \frac{A_{+}}{\Gamma} 
\frac{\partial}{\partial x} \right] \phi_0^{(0)}(X,\tau)= 
H_0(V) \phi_0^{(0)} - 
\frac{H_1(V)}{\Gamma} \frac{\partial \phi_0^{(0)}}{\partial X} +
\frac{H_2(V)}{2! \Gamma^2} \frac{\partial^2 \phi_0^{(0)}}{\partial X^2} +...,
\label{auto0}
\end{equation}
and represents a slowly decaying density inhomogeneity, 
$\phi_{0}^{0}(X,\tau_1)$,
with small (order $1/\Gamma$,  $1/\Gamma^2$, ...),  
{\it slaved} perturbations of
momentum, energy, etc..., whose shapes are
given by the successive derivatives of
the density distribution with respect to $X$. 
Similarly, the eigenfunction associated with $\mu=1$ has the explicit
representation
\begin{eqnarray}
\tilde P^{(1)}(X,V,\tau)= 
\exp(-\Gamma \tau) \left[ \left( H_1(V) \phi_0^{(1)} - 
\frac{H_2(V)}{\Gamma} \frac{\partial \phi_0^{(1)}}{\partial X} +
\frac{H_3(V)}{2! \Gamma^2} \frac{\partial^2 \phi_0^{(1)}}{\partial X^2} +...\right)+
\right. \nonumber 
\\
\left. + \frac{1}{\Gamma} \left(
H_0(V) \frac{\partial \phi_0^{(1)}}{\partial X} -
\frac{H_1(V)}{\Gamma} \frac{\partial^2 \phi_0^{(1)}}{\partial X^2} +...\right)
\right], 
\label{auto1}
\end{eqnarray}
where the first line in the r.h.s. has the interpretation of a current
inhomogeneity  $\phi_{0}^{(1)}(X,\tau/\Gamma)$, which {\it slaves} higher
order (energy,...) perturbations with decreasing amplitudes ($1/\Gamma$,...),
while the second line in the r.h.s. has the same structure of 
the $P^{(0)}(X,V,\tau)$
eigenfunction with amplitude 
$\phi_{0}^{(0)}=\Gamma^{-1} \partial_X \phi_{0}^{(1)}$,
and both terms have the fast decay of the exponential pre-factor. The physical
interpretation of such a combination is that an 
initially pure current fluctuation, described by
$H_1(V) \phi_1(X,0)$  would die very fast, as $\exp(-\Gamma \tau)$, 
but leaving behind a density fluctuation proportional to 
$\Gamma^{-1} \partial_X \phi_1^{(1)}(X,0)$, 
which would evolve with the {\it slow time} $\tau_1$. The particular
combination in (\ref{auto1}) is such that it completely cancels that 
remnant density
fluctuations, i.e. it orthogonalizes
$P^{(1)}(X,V,\tau)$ to $P^{(0)}(X,V,\tau)$,
and leaves a purely {\it fast} decaying form. 

  The structure of the higher order eigenvalues follows the same pattern, 
$P^{(2)}(X,V,\tau)$ is an energy fluctuation, decaying as  
$\exp(-2 \Gamma \tau)$,
but it has to contain diagonalizing terms proportional 
to $\Gamma^{-1} P^{(1)}$ and
to $\Gamma^{-2}  P^{(0)}$, to leave no slower remnant behind. With arbitrary 
choice of $\phi_0^{(\nu)}(X,0)$, for $\nu=0,1,2,...$, we may 
describe any initial
distribution of the ideal gas, whose time evolution would be given by the 
superposition of the decaying modes. These ``excited'' $\mu>0$ modes decay 
with a fast transient decay 
toward the only slowing decaying $\mu=0$ mode, which contains $\tau_1$ 
as the only relevant time scale. Such a separation between fast
decaying exponential modes, and a slow diffusive mode  
should be much more generic than the particular realization in 
the free ideal gas.
Indeed, it emanates from the structure 
of eq.(\ref{kramers0}), where the  heat-bath term
is associated with the diagonal operator 
of the form $\Gamma L_{FP}=-\Gamma \nu $, which contains a null 
matrix element ($\nu=0)$, 
while the remaining elements are proportional to $\Gamma$.
The non-diagonal contributions (given by the streaming 
terms for the ideal gas, and by collisions in general) 
are independent of 
$\Gamma$. In the limit
$\Gamma \gg 1$, the generic structure of the eigenfunctions, 
reflects the properties of the 
eigenfunctions of the $\Gamma L_{FP}$ operator, with corrections of
order $1/\Gamma$, that is combinations of exponential 
decays $\exp(-\nu \Gamma \tau)$
and slow functions, evolving with $\tau_1=\tau/\Gamma$, or slower. 
Therefore,
from an arbitrary initial condition, the system would have 
a fast transient decay
toward a {\it slow} mode, made of a density distribution, 
accompanied of
{\it slaved} current, energy, etc... fluctuations, 
with magnitude proportional to
inverse powers of $\Gamma$. In the next section we work 
out the leading contributions
of the collisions to that {\it slow} mode, taking into account their
non-linear character generates slower than $\tau_1$ times scales, as the slow
reaction to a slowly changing external force $F(X,\tau/\Gamma)$ would do.

\section{Multiple time-scale analysis}

How can we construct the equivalent
eigenfunction representation of Kramers' equation
for a system of interacting particles?
The method is provided by the multiple time-scale analysis, that
we shall discuss hereafter. The multiple time-scale method is designed to deal
with non uniformities in systems with more than a time scale.
It has been shown that a straightforward expansion of the Kramers
equation in powers of the small parameter $\Gamma^{-1}$ does
not lead to a uniformly valid result~\cite{Bocquet}.
In order to obtain a uniformly valid 
expansion, instead, one makes use of the presence of two different
time scales in the problem. The first scale, is fast and corresponds
to the time interval necessary to the velocities of the
particles to relax to configurations consistent with their thermal
equilibrium value. The second time scale is much longer and corresponds
to the time necessary to the positions of the particles
to assume their equilibrium configurations.

In the multiple time-scale analysis one determines the temporal evolution
of the distribution function $\tilde P(X,V,\tau)$ in the regime
$\Gamma^{-1}<<1$, by means of a perturbative method. 
In order to construct the solution one replaces the single
physical time scale, $\tau$, by a series of auxiliary time scales
($\tau_0,\tau_1,..,\tau_n$) which are related to the original variable
by the relations $\tau_n=\Gamma^{-n}\tau$. Also the original
time-dependent function, $\tilde P(X,V,\tau)$,  
is replaced by an auxiliary function,$\tilde P_a(X,V,\tau_0,\tau_1,..)$,  
which depends on the $\tau_n$, which are treated as independent variables.
Once the equations corresponding to the various orders have been 
determined, one returns to the original time variable and to the
original distribution.

One begins by replacing the time derivative
with respect to $\tau$ by a sum of partial derivatives:
\begin{equation}
\frac{\partial}{\partial \tau}=\frac{\partial}{\partial \tau_0}
+\frac{1}{\Gamma} \frac{\partial}{\partial \tau_1}
+\frac{1}{\Gamma^2} \frac{\partial}{\partial \tau_2}+..
\label{mult}
\end{equation}
First, the auxiliary function,$\tilde P_a(X,V,\tau_0,\tau_1,..)$
is expanded as a series of $\Gamma^{-1}$
\begin{equation} 
\tilde P_a(X,V,\tau_0,\tau_1,\tau_2,..)=
\sum_{s=0}^{\infty} \frac{1}{\Gamma^s} 
\tilde P_a^{(s)}(X,V,\tau_0,\tau_1,\tau_2,..).
\label{pn}
\end{equation}
Similarly, the collision operator is expanded as:
\begin{equation}
C_\alpha(X,\tau)=\sum_{s=0}^{\infty} \frac{1}{\Gamma^s} 
C_{s,\alpha}(X,\tau_0,\tau_1,\tau_2,..).
\label{cexp}
\end{equation}
Next, each term $P_a^{(s)}$ is projected over the functions $H_{\nu}$:
\begin{equation}
P^{(s)}_a(X,V,\tau_0,\tau_1,..)=\sum_{\nu=0}^{\infty}  
\psi_{s \nu}(X,\tau_0,\tau_1,\tau_2,..)H_{\nu}(V)
\label{phi}
\end{equation}
The term $C_{s,\alpha}$ represents the contribution 
of order $\Gamma^{-s}$ to $C_\alpha(X,\tau)$:
\begin{eqnarray}
C_{s,\alpha}(X,\tau)= \sum_{l+m=s}
\sum_{\mu,\nu} &&g_2(X,X+1)G_{\mu,\nu}^{\alpha}
\psi_{l\mu}(X,\tau)\psi_{m\nu}(X+1,\tau)\\\nonumber
&-&g_2(X,X-1)G_{\nu,\mu}^{\alpha}
\psi_{l\mu}(X,\tau)\psi_{m\nu}(X-1,\tau)\qquad . 
\label{coll_ab}
\end{eqnarray}

One substitutes, now, the time derivative~(\ref{mult})
and expressions~(\ref{pn})-(\ref{phi}) into eq. 
(\ref{kramers0}) and identifying terms 
of the same order in $\Gamma^{-1}$ in the equations
one obtains a hierarchy of relations
between the amplitudes $\psi_{s \nu}$. The advantage
of the method over the naive perturbation theory, is that 
secular divergences can be eliminated at each order of perturbation
theory and thus uniform convergence is achieved.

We show, now, how the method works. We 
substitute eqs.(\ref{mult})-(\ref{pn}) into eq. (\ref{kramers0})
and equating the coefficients of the same powers of $\Gamma$.
To order $\Gamma^0$ one finds:
\begin{equation}
L_{FP} \Bigr[\sum_\nu \psi_{0\nu}H_\nu\Bigr]=0
\label{g0}
\end{equation} 
and concludes that only the amplitude $\psi_{00}$ is non-zero. 

Next, we consider terms of order  $\Gamma^{-1}$ and write:
\begin{equation}
L_{FP} \Bigr[ \psi_{11}H_1+\psi_{12}H_2 \Bigr]=
\frac{\partial \psi_{00}}{\partial \tau_0}H_0+D_X\psi_{00}H_1
- C_{0,1}H_1- C_{0,2}H_2- C_{0,3}H_3
\label{g1}
\end{equation} 
having introduced, for notational convenience, 
$D_X\equiv(\partial_X-F(X,\tau))$.
Following the method of 
reference~\cite{Titulaer}, the amplitudes with $\nu=0$ and  $s>0$ are set
equal to zero.
Such a choice, although not unique is sufficient to eliminate secular terms,
i.e. terms containing a dependence on the slow time $\tau_0$.
By equating the coefficients multiplying the same $H_{\nu}$ 
we find that since:
\begin{equation}
\frac{\partial \psi_{00}}{\partial \tau_0}=0
\label{psi0ta}
\end{equation} 
the amplitude $\psi_{00}$ is not a function of  $\tau_0$.
Therefore, also the amplitude $\psi_{11}$, which is given by the relation
\begin{equation}
\psi_{11}=-D_X\psi_{00}+ C_{0,1},
\label{psi0t}
\end{equation} 
does not depend on $\tau_0$, being a functional of $\psi_{00}$, 
both through the linear operator $D_X$ and through the effective field
$C_{0,1}$, whose explicit form is given in section IV.
The remaining two amplitudes, instead,
vanish because to order $\Gamma^{-1}$ 
the self-consistent terms vanish, $C_{0,2}=0$ and $C_{0,3}=0$:
\begin{equation}
\psi_{12}=\frac{1}{2}C_{0,2}=0, 
\qquad \psi_{13}=\frac{1}{3}C_{0,3}=0
\label{psi0txx}
\end{equation}
In particular, the vanishing
of $C_{0,2}$ is a consequence of the traceless form (for an elastic
hard-rod system) of $G^2_{\mu,\nu}$ (see appendix A). 
A similar property yields $C_{0,3}=0$.

To order $\Gamma^{-2}$ we obtain the equation:
\begin{eqnarray}
L_{FP} &\Bigr[& \psi_{21}H_1+\psi_{22}H_2+
\psi_{23}H_3 \Bigr]
=\\\nonumber
&&\frac{\partial \psi_{11}}{\partial \tau_0}H_1
+\frac{\partial \psi_{00}}{\partial \tau_1}H_0+
D_X\psi_{11}H_2+\partial_X\psi_{11}H_0-C_{1,1}H_1-C_{1,2}H_2-C_{1,3}H_3
\nonumber
\end{eqnarray}
from which we obtain the  conditions:
\begin{equation}
\frac{\partial \psi_{00}}{\partial \tau_1}=
-\partial_X\psi_{11},
\label{psi0t1}
\end{equation}
and
\begin{equation}
\frac{\partial \psi_{11}}{\partial \tau_0}=
-\psi_{21}+ C_{1,1}.
\label{psi21a}
\end{equation}
Notice that, since the l.h.s. of eq.~(\ref{psi21a}) 
does not depend on $\tau_0$, as discussed after  eq.~(\ref{psi0t}), 
the r.h.s. must vanish.
Utilizing eqs. (\ref{psi0t}) and (\ref{psi0t1}) we write:
\begin{equation}
\frac{\partial \psi_{00}}{\partial \tau_1}=
\partial_X[D_X\psi_{00}- C_{0,1}]
\label{psi0t1b}
\end{equation}
By carrying on the procedure to  order $\Gamma^{-3}$ we 
obtain:

\begin{equation}
\frac{\partial \psi_{00}}{\partial \tau_2}=
-\partial_X\psi_{21}=- \partial_X C_{1,1}
\label{psi0t2}
\end{equation}
where we have used eq. (\ref{psi21a}) to eliminate $\psi_{21}$.

For the sake of completeness we write 
the third order correction  $\Gamma^{-3}$ and find:
\begin{equation}
\frac{\partial \psi_{00}}{\partial \tau_3}=
-\partial_X\Bigl[(\partial_X F)(D_X
\psi_{00}- C_{0,1})+C_{2,1}
-\frac{\partial C_{0,1}}{\partial \tau_1}
-\partial_X C_{1,2}\Bigl]
\label{psi0t3}
\end{equation}
The time derivative appearing in the r.h.s. can be expressed in terms
of spatial derivatives of the order parameter $\psi_{00}$
using eq. (\ref{psi0t1b}) and therefore could be computed. 

As a check of the method we have re-obtained perturbatively 
the exact solution in the ideal gas case. Moreover, 
equation~(\ref{psi0t3}) reduces to the modified
Smoluchowski diffusion equation in a potential 
obtained by Titulaer \cite{Titulaer}, who
showed that, in the case of independent particles in a parabolic
potential, it coincides 
with the exact solution up to order $\Gamma^{-3}$.

In the following, we shall truncate the expansion to second order.
Collecting together the various terms 
and employing eq.~(\ref{mult})  to eliminate the time variables
$\tau_0,\tau_1,\tau_2$ and restore the original time
variable $\tau$  we  obtain the evolution equation:
\begin{equation} 
\frac{\partial \psi_{00}}{\partial \tau}=
\frac{1}{\Gamma} \partial_X[D_X\psi_{00}- C_{0,1} -
\frac{1}{\Gamma}C_{1,1}]
\label{zero0}
\end{equation}
Clearly, the evolution equation~(\ref{zero0})
for the amplitude $\psi_{00}(X,\tau)$, representing the key result 
of the present paper,
has to be supplemented with a prescription for $C_{0,1}$
and $C_{1,1}$ which is given explicitly
in the next section. These terms represent collisions and involve
the density and current amplitude, $\psi_{00}$ and $\psi_{11}$,
respectively. However, the latter quantity
can be expressed by means of equation~(\ref{zero0})
as a functional of $\psi_{00}$. In this manner expression~(\ref{psi0t}) forms
a closed equation for the density profile. 

It is worth to remark that, while in the original 
hierarchy eq.~(\ref{brinkmanhiera}) the various amplitudes
were independent fields, the solution obtained in this section,
being, in fact, the generalization to interacting systems of the
zeroth eigenfunction of Kramers equation, imposes a constraint
on each of the $\nu>0$ components. We used this property as 
an internal check of the present extension to colliding particles.
Employing the constraint provided
by relations (\ref{psi0t}) and (\ref{psi0t1}) into 
the first equations of the hierarchy~(\ref{brinkmanhiera}),
we have verified that to order $\Gamma^{-2}$ indeed the method
provides a solution. In other words the solution
even in the presence of collisions can be represented only 
by the eigenfunction associated with the less negative eigenvalue.


\section{Evolution equation and its DDF limit}

Let us solve, in the case of interacting particles, the evolution 
equation (\ref{zero0}) for the amplitude $\psi_{00}(X,\tau)$, which corresponds
to the density fluctuation.
The collisional contributions of orders $\Gamma^{-1}$ and $\Gamma^{-2}$
are, respectively:
\begin{equation}
C_{0,1}= -\psi_{00}(X,\tau)\Bigl[g_2(X,X+1)
\psi_{00}(X+1,\tau)
-g_2(X,X-1)\psi_{00}(X-1,\tau)\Bigl]\Bigl\}
\end{equation}
and
\begin{eqnarray}
C_{1,1}= &&\frac{2}{\sqrt \pi}\psi_{00}(X,\tau)\Bigl[g_2(X,X+1)
\psi_{11}(X+1,\tau)
+g_2(X,X-1)\psi_{11}(X-1,\tau)\Bigl]\\-
&&\frac{2}{\sqrt \pi} \psi_{11}(X,\tau)\Bigl[g_2(X,X+1)
\psi_{00}(X+1,\tau)
+g_2(X,X-1)\psi_{00}(X-1,\tau)\Bigl]
\end{eqnarray}
where we have employed the matrix elements
$G^1_{0,0}=-1$, $G^1_{0,1}=2/\sqrt \pi$ and  $G_{1,0}=-G_{0,1}$
and relation
$\psi_{11}=-D_X\psi_{00}+ C_{0,1}$
to evaluate these expressions .
Notice that the self-consistent interaction term $C_{0,1}$ 
depends only on the amplitude $\psi_{00}(X,\tau)$ of the $H_0(V)$ component. 
It describes the contribution to the effective restoring force when the 
the velocity distribution of the colliding particles is Maxwellian.
The term $C_{1,1}$, instead, accounts for collisions between particles
whose velocity deviates from the equilibrium thermal distribution.
One may visualize, such a term by imagining the collision
as occurring between a thermalized particle, i.e. a particle with
zero average momentum, and a particle carrying momentum.
Indeed, the Langevin dynamics leading to the standard DDF equation 
describes only collision between perfectly thermalized particles.
 This can be seen, by using eq.~(\ref{zero0})
and neglecting the term $C_{1,1}$. One can recognize that the following 
equation
\begin{eqnarray}
\frac{\partial \psi_{00}(X,\tau)}{\partial \tau}&=&\frac{1}{\Gamma}
\frac{\partial }{\partial X}
\Bigl \{\frac{\partial \psi_{00}(X,\tau)}{\partial X}
-F(X,\tau)\psi_{00}(X,\tau)\\
\nonumber
&+&\psi_{00}(X,\tau)\Bigl[g_2(X,X+1)
\psi_{00}(X+1,\tau)
-g_2(X,X-1)\psi_{00}(X-1,\tau)\Bigl]\Bigl \}
\end{eqnarray}
represents the governing equation of the DDF method, 
expressed in dimensionless guise.

It is also worth to comment the fact that the short-range and the long-range
contributions to the dynamics, contained
in $C_{s,\nu}(X,\tau)$ and $F(X,\tau)$ respectively, 
do not appear on equal footing.
This state of affairs is encountered also when studying the
equilibrium properties of liquids and was first recognized by van der Waals.
In the present dynamical approach we see that the 
difference originates in the fact that in the hard core term, $K(X,V,\tau)$,
the velocity dependence of the distribution function $\tilde P(X,V,\tau)$
does not factorize as in the molecular field term. The effect of 
hard-core collision depends not only on the amplitude
of the Maxwellian component of the velocity distribution, 
but on the full velocity
distribution. Therefore, as far as the system is not fully thermalized
we observe a force which has not counterpart in equilibrium systems.  
However, as the system relaxes the term $C_{1,1}$
tends to zero, because its amplitude depends on the current
$\psi_{11}(X,\tau)=-D_X\psi_{00}(X,\tau)+ C_{0,1}(X,\tau)$ 
which vanishes at equilibrium. 


\subsection{Linear analysis}

The following simple example may give an idea of the
role of the corrections to the DDF equation.
We compare the analytical results of our theory
with those obtained by computer simulations of the model described by eq.
(\ref{kramers}) for an ensemble
of $N$ hard particles stochastically driven, in a periodic box of
size $L$~\cite{Cecconi}. 
Particle positions and
velocities within two consecutive collisions are updated according to
a second order discretization scheme for the
dynamics eq.~(\ref{kramers}). Averages over $10^4$ realizations of the
noise were taken.

We perform the analysis of the
evolution of a small initial perturbation $\Delta \rho(X,\tau)\sigma=
(\psi_{00}(X,\tau)-\rho_0\sigma)$ and show
that while the DDF predicts that the relaxation depends
only on the time scale $\tau/\Gamma$, hence is universal,
the present theory leads to a violation of this scaling.

In the limit of vanishing perturbations, each Fourier component evolves 
independently, and decays to zero exponentially~\cite{Penna}. 

The characteristic relaxation
time can be ascertained by substituting in eq.
(\ref{zero0})
the trial solution
$\psi_{00}(X,\tau)=\rho_0\sigma+ A(\tau) \cos (kX)$
and keeping only linear terms. The resulting equation reads
\begin{equation}
\frac{\partial A(\tau)}{\partial \tau}=-  
\frac{\alpha(K)}{\Gamma}  A(\tau)+ {\cal{O}}(A^2),
\label{sol_linear} 
\end{equation}
and has an exponential solution with a wave-length dependent decay time
\begin{equation}
\frac{\alpha(K)}{\Gamma}=
\frac{1}{\Gamma}\frac{K^2}{S(K)}(1-\epsilon(K))
\label{at} 
\end{equation}
where $S(K)$ is the hard-rod equilibrium structure factor
\begin{equation}
S(K)=\Bigl[1 + 2 p_o\sigma \frac{sin(K)}{K} + 
4 (p_o\sigma)^2 
\frac{\sin^2\left( \frac{K}{2} \right)}{K^2}\Bigl]^{-1}, 
\label{DDF_decay} 
\end{equation}
with $p_o=\rho_0/(1-\rho_0\sigma)$, 
being the pressure of the uniform 1D fluid divided by $k_B T$.
In physical units the decay time associated with the
wave-vector $q=K/\sigma$ reads $D q^2 (1-\epsilon) /S(q\sigma)$,
where for the self-diffusion coefficient, $D$, we use the result for isolated 
Brownian spheres, given by the Stokes-Einstein equation:
\begin{equation}
D=\frac{k_B T}{6\pi\eta\sigma}
\end{equation}
$\eta$  being the viscosity of the suspending fluid.
The correction to the DDF result, $\epsilon(K)$, appearing in eq.(\ref{at}),
reads
\begin{equation}
\epsilon(K)=\frac{8 p_o\sigma}{\Gamma \sqrt{\pi}}  
\sin^2\left( \frac{K}{2}\right)=
\frac{4 \omega_E}{\gamma}\sin^2\left( \frac{K}{2}\right)
\label{decay} 
\end{equation}
and depends on the ratio
$\frac{8 p_o\sigma}{\Gamma \sqrt{\pi}}=4\frac{\omega_E}{\gamma}$ 
between the Enskog collision frequency and the heat-bath characteristic 
frequency. 
For a uniform one dimensional hard rod system the collision frequency is
\begin{equation}
\omega_{E}= 2 \frac{v_T}{\sqrt{\pi}}p_o=2 \sqrt{\frac{k_B T}{m \pi}}p_o
\label{decay2}
\end{equation}


The prediction of the present theory for the variation of the relaxation time
with respect to the wave-vector of the perturbation 
is shown in Fig.1  and compared with the DDF result.

 Figure 2 illustrates the comparison between 
the theoretical predictions and numerical simulations in the case
of an initial sinusoidal perturbation of period $K=2 \pi/1.4$ and several
values of the dimensionless friction $\Gamma$. Instead of the data collapse
predicted by DDF we observe deviations in the short-time regime. Only
for large value of $\Gamma$, i.e. in the over-damped limit, we recover
universality. 

In particular, one observes a slower relaxation of density fluctuations.
The larger the collision frequency, the slower the decay.
In other words the theory predicts that, at fixed $\Gamma$,
collisions render the relaxation process slower. What is
the physical origin of this slowing down?
One can think that a current of momentum  can occur either
via a particle displacement, i.e. a density change, or through 
collisional transfer.
However, in the latter case the momentum can travel
a distance $\sigma$ without paying any price to the frictional
force $-m\gamma v$. Such a mechanism renders this 
relaxation ``channel'' slower.
The two type of relaxation processes are sketched in Fig.3.
The first process dominates when the system is close to equilibrium
when the velocity distribution is well described by a Maxwellian.
In the second collision process, which 
is more relevant far from equilibrium, the two distributions have the 
same temperature, but not the same momentum.

Two facts are worth to mention: a) the correction has a kinetic origin as
can be seen from the presence of the mass in the 
last member of eq.~(\ref{decay2}). When $m\to\infty$ the
correction vanishes, being the inertial effect negligible. 
On the contrary, in the case of over-damped dynamics, $\epsilon(K)\to 0$,
the mass does not appear explicitly in the diffusion coefficient $D$
and only geometrical factors such as $S(K)$ play a role.
Secondly,
the correction increases as the particle size, $\sigma$, increases.

\section{Conclusions}

We have considered the non equilibrium colloidal dynamics of a 
system of hard rods of mass $m$ driven by a uniform heat bath.
The evolution depends on the non dimensional parameter $\Gamma^{-1}$,
proportional to the time span occurring to the velocity distribution
to reach its equilibrium value, and on the packing fraction.
This evolution is described by a Kramers equation for the phase-space
density $P(X,V,\tau)$ supplemented by a collision term, treated within
the Revised Enskog Theory. 
Since the momentum degrees of freedom equilibrate much faster than the
positional degrees of freedom it is reasonable to look for a description
which contains only the latter variables.
By employing the multiple-time scale method 
we have performed the $\Gamma^{-1}$ expansion
of Kramers-Enskog equation and obtained a modified
Smoluchowski-Enskog equation for the density field.
We found that the collision term gives a non-local
coupling between density, momentum and energy fluctuations.
However, the density field slaves the remaining fields.
To lowest order in $\Gamma^{-1}$ 
the present method yields the same evolution equation for the 
density as the one obtained within the DDF
approach. 
The present derivation does not require
the existence of any equilibrium density functional, 
but is based on kinetic theory arguments. Therefore, it
can be applied to generic non-equilibrium systems, where the RET closure
of the evolution equation for the phase-space distribution   
is physically sound.
However,  containing as a key ingredient 
the same equilibrium pair correlation as the DDF, the matching
between the two methods is not too surprising.

As discussed by Archer and Evans~\cite{Archer} if the thermal equilibration
occurs mainly via the solvent the deviations from 
the DDF should be negligible.
Nevertheless, for atomic fluids the harshly repulsive potential might
concur appreciably to the relaxation process and lead to significant
effects which are beyond the limits of the DDF approach.

Besides reproducing known results the present derivation
provides systematic corrections to the DDF equation accounting for
the deviation of the velocity distribution from the Maxwellian.
Hence, it can describe situations very far from thermodynamic equilibrium
or even situations where a steady, but non-equilibrium state exists.

The present method quite naturally lends itself to the following 
future applications and extensions:
a) hard core systems whose spatial
dimensionality is larger than one, 
b) systems of particles experiencing
inelastic collisions, such as granular gases, 
where free energy functional approaches
are not applicable \cite{Ernst} and the RET closure provides a
valid alternative,
c) systems having a non-uniform temperature 
profile~\cite{Matsuo,Vankampen2} where the
standard isothermal DDF approach cannot be applied,
d) inclusion of higher order corrections in the inverse friction
expansion $\Gamma^{-1}$ accounting for currents associated with higher
moments of the velocity distribution.

\section{Acknowledgments}
We wish to thank F.Cecconi for providing the 
numerical code on which our simulations are based
and G.Costantini for help with the plots.
This work was supported by the Ministerio de Educaci\'on
y Ciencia of Spain grant N. SAB2003-0207.
U.M.B.M. thanks the Departamento de Fisica Teorica de la Materia
Condensada of the Universidad Autonoma de Madrid for hospitality.

\appendix
\section{Collision integrals}

We consider explicitly
the first three coefficients $C_{\alpha}$ featuring in the series expansion
(\ref{expansion3}):
\begin{equation}
C_{\alpha}(X,\tau)=\int_{-\infty}^{\infty}dV 
\mu_{\alpha}(V)K(X,V,\tau)
\label{coll_adef}
\end{equation}
with $\mu_0=1$,  $\mu_1=V$ and $\mu_2=(V^2-1)/2$.
Using the definition of $K$ given by eq.(\ref{jcollis1}) one finds
the expressions:
\begin{eqnarray}
C_{\alpha}(X,\tau)&=&g_2(X,X+1)\Bigl\{
\int_{-\infty}^{\infty}dV \mu_{\alpha}(V)\\
\nonumber
&&\Bigl[\int_{-\infty}^{0}du u 
 \tilde P(X,V,\tau)\tilde P(X+1,u+V,\tau)+
\int_0^{\infty} du u 
\tilde P(X,u+V,\tau)\tilde P(X+1,V,\tau)\Bigl]\Bigl \}\\\nonumber
&&-g_2(X,X-1)\Bigl\{
\int_{-\infty}^{\infty}dV \mu_{\alpha}(V)\\\nonumber
&&\Bigl[\int_{-\infty}^0 du u 
\tilde P(X,u+V,\tau)\tilde P(X-1,V,\tau)+
\int_{0}^{\infty}du u 
\tilde P(X,V,\tau)\tilde P(X-1,u+V,\tau)\Bigl]\Bigl \}
\label{coll_alongbb}
\end{eqnarray}
After substituting expansion~(\ref{expansion}) 
into eq.~(A2) and integrating over velocities
one  obtains:
\begin{equation}
C_{\alpha}(X,\tau)=g_2(X,X+1)\sum_{\mu,\nu}G_{\mu,\nu}^{\alpha}
\phi_{\mu}(X,\tau)\phi_{\nu}(X+1,\tau)
-g_2(X,X-1)\sum_{\mu,\nu}G_{\nu,\mu}^{\alpha}
\phi_{\mu}(X,\tau)\phi_{\nu}(X-1,\tau)
\label{coll_a}
\end{equation}
where the matrix elements $G_{\mu,\nu}^{\alpha}$ are given by
\begin{equation}
G_{\mu,\nu}^{\alpha}=
\int_{-\infty}^{\infty}dV \mu_{\alpha}(V)\Bigl[
\int_{-\infty}^0 du u H_{\mu}(V)H_{\nu}(u+V)
+\int_{0}^{\infty}du u H_{\nu}(V)H_{\mu}(u+V)\Bigl]
\label{gmunu}
\end{equation}
The integral $C_0(X,\tau)=0$ is zero, as required
by the conservation of the number of particles during
a collision, and indeed  all $G_{\mu,\nu}^{0}=0$ vanish.
The explicit form of the $G_{\mu,\nu}^{\alpha}$ 
for $\mu,\nu=0,1,2$ and $\alpha=1,2$ are given 
by the following matrices:

\[ G_{\mu,\nu}^{1} = \left| 
\begin{array}{ccc}
-1 & \frac{2}{\sqrt \pi} & -1  \\
-\frac{2}{\sqrt \pi} & 1 &-\frac{1}{\sqrt{ \pi}}  \\
-1 & \frac{1}{\sqrt{\pi}} & 0 \\
\end{array}
\right|.\] 

and 
\[ G_{\mu,\nu}^{2} = \left| 
\begin{array}{ccc}
0 & -\frac{1}{2} & \frac{2}{\sqrt{ \pi}}  \\
-\frac{1}{2} & 0 &\frac{1}{2}  \\
-\frac{2}{\sqrt {\pi}} & \frac{1}{{ 2}} & 0 \\
\end{array}
\right|.\]

\section{A useful identity}
We prove hereafter that 
the collision kernel $K(X,V,\tau)$ 
and thus the expansion coefficients $C_n(X,\tau)$ can be expressed as
divergences. To this purpose we employ the following identity~\cite{Brey}:
\begin{eqnarray}
S(X,X+Y)-S(X-Y,X)&=&\int_0^1 dz\frac{\partial}{\partial z}
S(X-(1-z)Y,X + z Y)\\
\nonumber
&=&\frac{\partial}{\partial X}\int_0^1 dz S(X-(1-z)Y,X + z Y)
\label{iden}
\end{eqnarray}
and identify
\begin{eqnarray}
S(X,X+Y)&=&-g_2(X,X+Y)\int_{-\infty}^{\infty} dV_2  (V_1-V_2)\\
\nonumber
&&\Bigl\{\Theta(V_1-V_2)P(X,V_1,\tau) P(X+Y,V_2,\tau)
+\Theta(V_2-V_1)P(X+Y,V_1,\tau)P(X,V_2,\tau)\Bigl\}
\label{iden19}
\end{eqnarray}
and setting $Y=1$ rewrite eq.(\ref{jcollis1}) with the help of 
eq.(\ref{iden}) as:

\begin{eqnarray}
&&K(X,V_1,\tau)=-\frac{\partial}{\partial X}\int_0^1 dz
g_2(X-(1-z)Y,X+zY)\int_{-\infty}^{\infty} dV_2 (V_1-V_2)\times\\ \nonumber
&&\Bigl\{\Theta(V_1-V_2)P(X-(1-z)Y,V_1,\tau) P(X+zY,V_2,\tau)\\
\nonumber
&+&\Theta(V_2-V_1)P(X+zY,V_1,\tau)P(X-(1-z)Y,V_2,\tau)\Bigl\}
\label{diverg}
\end{eqnarray}



\section*{FIGURE CAPTIONS}
 
{\large{\bf Fig.1}}
\noindent
Exponential relaxation time corresponding to a
sinusoidal density modulation of
wave-vector K. The uniform background is $\rho_0\sigma=0.6$. 
The dashed line
represents the DDF results, while the continuous line represents the result
of the present theory with $\Gamma=10$ and the dots the non interacting 
system.

{\large{\bf Fig.2}}
\noindent
Initial stage of
the temporal evolution of the
amplitude decay of a sinusoidal modulation for different values
of the dimensionless parameter $\Gamma$. The DDF theory would predict a 
collapse of the data when using the time scale $\tau/\Gamma$, whereas the
present theory predicts a $\Gamma$ dependence. The circles represent
the numerical results relative to $\Gamma=5$, square symbols refer to
$\Gamma=7$, diamonds to $\Gamma=10$, upper triangles $\Gamma=20$ and
down triangles to $\Gamma=100$.
In the inset we report the same data using the original time scale $\tau$.
The straight lines correspond
to the values of the relaxation time predicted by the linearized solution.

{\large{\bf Fig.3}}
\noindent
Sketch of the two different collision processes contributing to
the term $C_{0,1}$ (process I) and to the term $C_{1,1}$ (process II).

\newpage

\includegraphics[clip=true,width=15.cm, keepaspectratio]
{fig1.eps}

\LARGE{FIGURE 1}
\newpage
\includegraphics[clip=true,width=15.cm,keepaspectratio,angle=0]
{fig2.eps}

\LARGE{FIGURE 2}
\newpage
\includegraphics[clip=true,width=15.cm, keepaspectratio,angle=-90]
{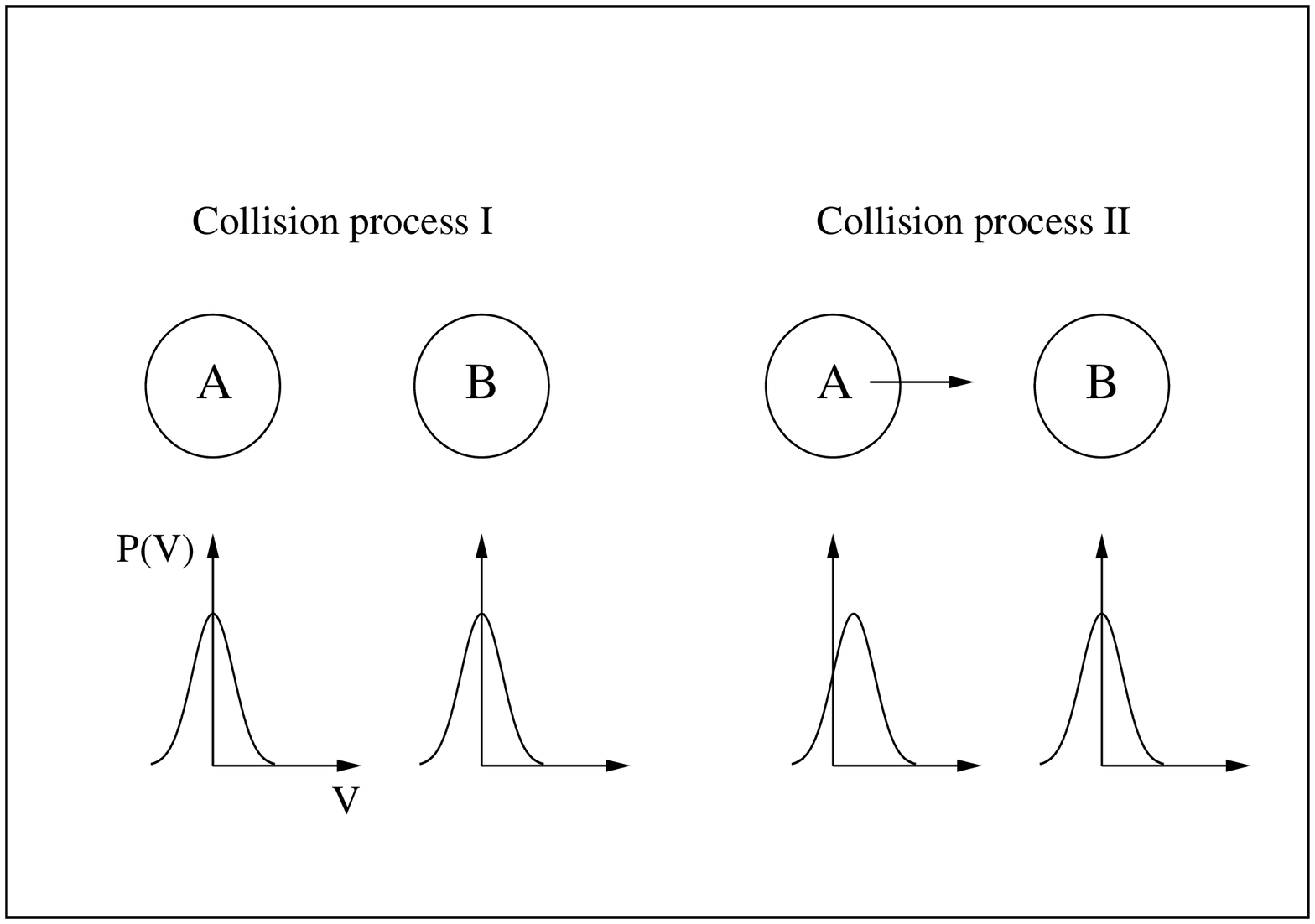}

\LARGE{FIGURE 3}

\end{document}